\documentclass[11pt]{article}
\usepackage{amsfonts, latexsym, amssymb}
\makeatletter
\def \underset#1#2{\mathrel{\mathop{#2}\limits_{#1}}}

\@addtoreset{equation}{section}

\def\bm{\left(\begin{array}{cc}}
\def\em{\end{array}\right)}

\newtheorem{remark}{Remark}[section]

\newtheorem{proposition}{Proposition}[section]
\def\text#1{\mbox{#1}}

\def\lim{\mathop{\mbox{lim}}\limits}

\makeatother
\begin{document}
\title{Non-Schlesinger Deformations
of Ordinary Differential Equations with Rational Coefficients}
\author{A.~V.~Kitaev
\thanks{E-mail: kitaev@pdmi.ras.ru}\\
Steklov Mathematical Institute, Fontanka 27, St.Petersburg, 191011, Russia\\
and\\
Department of Pure Mathematics\\
University of Adelaide, Adelaide, SA 5005, Australia}

\date{June 15,2000}
\maketitle

\begin{abstract}
We consider deformations of $2\times2$
and $3\times3$ matrix linear ODEs with rational coefficients 
with respect to singular points 
of Fuchsian type which don't satisfy the well-known system of 
Schlesinger equations (or its natural generalization). 
Some general statements concerning reducibility of such deformations 
for $2\times2$ ODEs are proved. 
An explicit example of the general non-Schlesinger deformation
of $2\times2$-matrix ODE of the
Fuchsian type with $4$ singular points is constructed and
application of such deformations to the construction of special solutions 
of the corresponding Schlesinger systems is discussed. 
Some examples of isomonodromy and non-isomonodromy deformations
of $3\times3$ matrix ODEs are considered. The latter arise as 
the compatibility conditions with linear ODEs with non-singlevalued 
coefficients.
\vspace{24pt}\\
%\end{abstract}
\vspace{24pt}\\
Short title: Non-Schlesinger Isomonodromy Deformations
\end{abstract}
\maketitle
\newpage
\section{Introduction}
It is well-known that isomonodromy deformations of the Fuchsian matrix ODE,
\begin{equation}
 \label{eq:A}
\frac{d\Psi}{d\lambda}=\left(\frac{A_0}{\lambda}+\frac{A_1}{\lambda-1}
+\frac{A_t}{\lambda-t}\right)\Psi,
\end{equation}
where $A_\infty=A_0+A_1+A_t$ is normalized to be independent of $t$,
are governed by the system of Schlesinger equations \cite{S},
\begin{equation}
 \label{eq:S}
\frac{dA_0}{dt}=\frac1t[A_t,\,A_0],\quad
\frac{dA_1}{dt}=\frac1{t-1}[A_t,\,A_1],\quad
\frac{dA_t}{dt}=[\frac1tA_0+\frac1{t-1}A_1,\,A_t].
\end{equation}  
This system is the compatibility condition of Eq.~(\ref{eq:A}) with the
following ODE,
\begin{equation}
 \label{eq:T}
\frac{d\Psi}{dt}=-\frac{A_t}{\lambda-t}\Psi.
\end{equation}
Throughout the paper we consider matrix ODEs with respect to independent
variable $\lambda$ whose coefficients are rational functions of
$\lambda$ with at least one Fuchsian singularity at $\lambda=t$. We
call deformations of such ODEs Schlesinger (with respect to $t$) if
these ODEs are compatible, with Eq.~(\ref{eq:T}). This is a natural
generalization of the notion of Schlesinger deformations of the 
Fuchsian ODEs. It is clear that the Schlesinger deformations
are the simplest isomonodromy deformations. On the other
hand, as it follows from the work by Malgrange \cite{M} any isomonodromy
deformation of the Fuchsian system could be described {\it up to
isomorphism} by the Schlesinger deformations. The purpose of this work
is to understand better what does ``up to isomorphism'' mean from the
point of view of the theory of integrable systems and special functions 
of the isomonodromy type \cite{K}. More precisely, the question is whether
this isomorphism can be always constructed explicitly, or by other words 
can we obtain any new integrable systems or special functions by considering 
non-Schlesinger deformations of ODEs with rational coefficients?   

The results presented in Sections \ref{sec:2}
and \ref{sec:3} give a negative answer on this question
in the case of the isomonodromy deformations: any integrable
system, which describes non-Schlesinger isomonodromy deformations of any 
Fuchsian ODE (or weak non-Schlesinger (see Section \ref{sec:2}) isomonodromy 
deformation for non-Fuchsian ODEs with rational coefficients) 
can be mapped explicitly by a proper Schlesinger {\it transformation} into 
some integrable system describing deformation of the Schlesinger type.  
This statement can be 
reformulated in the following way: there are no new transcendental 
functions defined via non-Schlesinger isomonodromy deformations of the
Fuchsian ODEs. The sketch of this rather simple proof is given at the 
end of Section \ref{sec:3}. It was communicated to me by A.~A.~Bolibruch 
for the case of the Fuchsian ODEs few weeks after the workshop where the 
results of this work were reported. Moreover, he pointed out to me his 
recent work \cite{B}, where he also studied non-Schlesinger deformations 
of the Fuchsian ODEs but looking at them from a different angle: the 
isomonodromy confluences of the Fuchsian singularities. However, 
A.~A.Bolibruch did not study the reducibility of the ODEs subject to 
non-Schlesinger deformations which I address, in particular, in this
work. The latter means further simplification of these ODEs.
Moreover, such reducibility does not follow from the proof mentioned
above and has a remarkable consequence for construction of the explicit
solutions of the Painlev\'e and higher Painlev\'e equations, and 
(generalized) Garnier type systems.

In Section \ref{sec:4} some simplest examples 
of non-isomonodromy deformations of $3\times3$ matrix ODEs, which are, 
of course, non-Schlesinger, are presented.
Here for simplicity we consider non-Fuchsian type ODE since one of our
tasks here is to show that non-isomonodromy deformations, which
can be described via compatibility conditions (generalized Schlesinger
equations), really exist. 
For the Fuchsian ODEs one can consider analogous examples.
In our examples, presented in Section \ref{sec:4}, appear only elementary 
and the Painlev\'e type functions. The latter non-evident fact is 
established (also after the workshop) by C.~M.~Cosgrove. 
However, to the best of my knowledge no general statements concerning 
reducibility of such deformations to the ones of the isomonodromy type 
are known. So that further interesting studies of such deformations are 
anticipated.

\section{Non-Schlesinger Isomonodromy Deformations of $2\times2$
Matrix ODEs}
 \label{sec:2}
We begin our study with the special case of Eq.~(\ref{eq:A}) when
$A_k\in sl_2(\mathbb C)$ for $k=0,\,1,\,t,\,\infty$. Let us
suppose that $A_\infty=-\frac{\theta_\infty}2\sigma_3,\;\theta_\infty\neq0$, 
which is in fact (excluding one exceptional solvable case) 
also a normalization rather than a restriction on $A_k$.
In this case the system of Schlesinger equations (\ref{eq:S}) 
is equivalent to the sixth Painlev\'e equation \cite{JM}. 

Eq.~(\ref{eq:T}) is a sufficient condition that monodromy matrices 
of a fundamental solution of Eq.~(\ref{eq:A}) are independent of $t$.
In general situation this condition is also necessary, but there are some
special exceptional cases, which are discussed here. A possibility of
such deformations resulted from the non-uniqueness of solubility of the
inverse monodromy problem for some special monodromy matrices.

We begin with a simplest generalization of Eq.~(\ref{eq:T}),
\begin{equation}
 \label{eq:T1}
\frac{d\Psi}{dt}=-\frac{A_t+\Lambda}{\lambda-t}\Psi.
\end{equation}
Compatibility condition of Eqs.~(\ref{eq:A}) and (\ref{eq:T1})
reads,
\begin{eqnarray}
 \label{eq:L}
&[\Lambda,\,A_t]=\Lambda,&\\
&\frac{dA_0}{dt}=\frac1t[A_t+\Lambda,\,A_0],\quad
\frac{dA_1}{dt}=\frac1{t-1}[A_t+\Lambda,\,A_1],\quad
\frac{dA_t}{dt}=[\frac1tA_0+\frac1{t-1}A_1,\,A_t+\Lambda].&\quad
\label{eq:AL}
\end{eqnarray}
Eq.~(\ref{eq:L}) implies:
${\rm tr}\Lambda=0,\;\Lambda^2=0,\;{\rm and}\;\pm1/2$ 
are the eigenvalues of $A_t$. 
Denote
$$
A_t=\left(\begin{array}{cc}
a_t&b_t\\
c_t&-a_t
\end{array}\right),
$$
then solution of Eq.~(\ref{eq:L}) can be written as follows
\begin{equation}
 \label{eq:LAMBDA1}
\Lambda=\mu\left(\begin{array}{cc}
-(a_t+1/2)b_t&-b_t^2\\
(a_t+1/2)^2&(a_t+1/2)b_t
\end{array}\right)\quad {\rm for}\quad a_t\neq-1/2,
\end{equation}
and
\begin{equation}
 \label{eq:LAMBDA2}
\Lambda=\mu\left(\begin{array}{cc}
c_t&1\\
-c_t^2&-c_t
\end{array}\right)\quad{\rm for}\quad a_t=-1/2,
\end{equation}
where $\mu=\mu(t)$ is an arbitrary function of $t$.\\
Consider now asymptotic expansion of a fundamental solution of 
Eq.~(\ref{eq:A}) as $\lambda\to t$,
\begin{equation}
 \label{eq:psi_at_t}
\Psi=\sum\limits_{k=0}^\infty\psi_k(\lambda-t)^k
(\lambda-t)^{\sigma_3/2}
\left(\begin{array}{cc}
1&\kappa\ln(\lambda-t)\\
0&1
\end{array}\right)
C,\quad \det C=\det\psi_0=1,
\end{equation}
where the parameter $\kappa=0$ or $1$. One finds
$$
A_t=\frac12\psi_0\sigma_3\psi_0^{-1},\quad
\Lambda=\left(C_{21}'C_{22}-C_{22}'C_{21}\right)\psi_0
\left(\begin{array}{cc}
0&0\\
1&0
\end{array}\right)\psi_0^{-1},
$$
where $C_{21}$ and $C_{22}$ are the corresponding matrix
elements of $C$ and the primes denotes differentiation on $t$.
The monodromy matrix of $\Psi$ corresponding to any loop, with only one
singular point $\lambda=t$ inside, reads,
$$
M_t=-I-2\pi i\kappa
\left(\begin{array}{cc}
C_{21}C_{22}&C_{22}^2\\
-C_{21}^2&-C_{21}C_{22}
\end{array}\right).
$$ 
If $\kappa\neq0$ then the isomonodromy condition, 
$\frac{d}{dt}M_t=0$, implies
$C_{21}'C_{22}-C_{22}'C_{21}=0$ and therefore $\Lambda=0$.
Thus, in the case of $2\times2$ matrices non-Schlesinger
deformations of the type (\ref{eq:T1}) exist if and only
if $\kappa=0$. This condition can be written in terms of
Eq.~(\ref{eq:A}) as follows
\begin{equation}
\mathrm{tr}\left((\frac1tA_0+\frac1{t-1}A_1)(A_t+\frac12\sigma_3)\right)=0.
\end{equation}
Consider explicit construction of this deformation.
Taking into account the integrals of the Schlesinger system (\ref{eq:S}),
$$
A_t=-A_0-A_1-\frac{\theta_\infty}2\sigma_3,\quad\mathrm{tr}A_k=0,
\quad\det A_k=-\frac{\theta_k^2}4,
$$
where $\theta_t=1$ and $\theta_k,\;k=0,\,1,\,\infty$ are the constants of 
integration considered as parameters;
we see that the most general non-Schlesinger deformation of the type
(\ref{eq:T1}) depends on the function $\mu(t)$ (see Eqs.~(\ref{eq:LAMBDA1}) 
and (\ref{eq:LAMBDA2})) and five parameters (constants of integration
of Eqs.~(\ref{eq:AL})):  
$\theta_0,\;\theta_1,\theta_\infty$ and $c_1$, $c_2$.  
One of them can be introduced via the gauge transformation,
\begin{equation}
 \label{eq:gauge}
A_k\longrightarrow c_2^{\sigma_3}A_kc_2^{-\sigma_3}.
\end{equation}
Due to the absence of the $\log$-term in the expansion Eq.~(\ref{eq:psi_at_t}),
the singularity at $\lambda=t$ in Eq.~(\ref{eq:A}) is removable via a 
proper Schlesinger transformation. Let us describe this construction in detail.
Consider the hypergeometric equation which can be written in the matrix
form as follows,
\begin{equation}
 \label{eq:hyper}
\frac{d\Phi}{d\lambda}=\left(
\frac A{\lambda}-
\frac{\frac{\theta_\infty+1}2\sigma_3+A}{\lambda-1}\right)\Phi,\quad
A=\left(\begin{array}{cc}
-\frac{\theta_0^2-\theta_1^2+(\theta_\infty+1)^2}{4(\theta_\infty+1)}&
\frac{\theta_0^2-(\theta_\infty+1-\theta_1)^2}{4(\theta_\infty+1)}\\
-\frac{\theta_0^2-(\theta_\infty+1+\theta_1)^2}{4(\theta_\infty+1)}&
\frac{\theta_0^2-\theta_1^2+(\theta_\infty+1)^2}{4(\theta_\infty+1)}
\end{array}\right).
\end{equation}
Note that $\det A=-\frac{\theta_0^2}4$ and 
$\det(\frac{\theta_\infty+1}2\sigma_3+A)=-\frac{\theta_1^2}4$.
An explicit formula for the fundamental solution $\Phi=\Phi(\lambda)$ of 
this equation in terms of the Gau{\ss} hypergeometric functions
can be found for example in \cite{J}. In the following we 
denote as $\Phi$ any fundamental solution of Eq.~(\ref{eq:hyper})
which is independent of $t$. A fundamental solution of the
system (\ref{eq:A}), (\ref{eq:T}) which corresponds to the general
(modulo gauge transformation (\ref{eq:gauge})) solution of the
system (\ref{eq:T1}), (\ref{eq:A}) (for $\theta_t=1$) can be written as 
follows,
\begin{eqnarray}
 \label{eq:SCH}
\Psi&=&D^{-1}R^{-1}(\lambda,t)\Phi,\quad
D=\left(\begin{array}{cc}
1&1\\
0&\kappa_1
\end{array}\right),
\kappa_1=\frac{4\theta_\infty(\theta_\infty+1)}
{(\theta_\infty+1-\theta_1)^2-\theta_0^2},\\
R(\lambda,t)&=&\left(\left(\begin{array}{cc}
0&0\\
0&1
\end{array}\right)
\sqrt{\lambda-t}+
\left(\begin{array}{cc}
1&0\\
c_1(t)&0
\end{array}\right)
\frac1{\sqrt{\lambda-t}}\right),\quad \det R(\lambda,t)=1,
\label{eq:SCH2}
\end{eqnarray}
where $c_1(t)$ is an arbitrary function of $t$ and for both square roots
in Eq.~(\ref{eq:SCH2}) should be chosen the same branch. 
The corresponding solution of the system (\ref{eq:L}), (\ref{eq:AL}) reads:
\begin{equation}
 \label{sys:Ansc}
\begin{array}{cl}
 \label{sys:nsc}
A_0=D^{-1}R^{-1}(0,t)ADR(0,t),&
A_1=-D^{-1}R^{-1}(1,t)\!\left(\!\frac{\theta_\infty+1}2\sigma_3+A\!\right)
\!DR(1,t),\\
A_t=D^{-1}
\left(\begin{array}{cc}
1/2&0\\
r(t)&-1/2
\end{array}\right)
D,&
\Lambda\;\,=\;\,D^{-1}
\left(\begin{array}{cc}
0&0\\
\frac{dc_1(t)}{dt}-r(t)&0
\end{array}\right)
D,
\end{array}
\end{equation}
where
\begin{eqnarray}
&r(t)=\frac{(\theta_0^2-(\theta_\infty+1-\theta_1)^2)c_1^2(t)+
(4t(\theta_\infty+1)^2-2((\theta_\infty+1)^2+\theta_0^2-\theta_1^2))c_1(t)+
\theta_0^2-(\theta_\infty+1+\theta_1)^2}{4t(t-1)(\theta_\infty+1)}.&
\end{eqnarray}
The function $c_1(t)$ depends on the function $\mu(t)$ and
the parameter $c_1$ which is introduced in the paragraph above 
Eq.~(\ref{eq:gauge}). Actually, if the function $\mu(t)$ is given,
as one of the coefficients in the system (\ref{eq:L}), (\ref{eq:AL}), then 
\begin{equation}
 \label{eq:c1}
\mu(t)=\frac{\kappa_1(c_1'(t)-r(t))}{(r(t)-\kappa_1)^2}.
\end{equation}
So that the function $c_1(t)$ is the general solution of the
differential equation (\ref{eq:c1}) and, therefore, depends on the 
constant of integration $c_1$. 

At this stage it is worth to notice an application of the non-Schlesinger
deformations to the construction of one-parameter families of solutions 
of the sixth Painlev\'e equation. As is mentioned above, the sixth 
Painlev\'e equation corresponds to the Schlesinger deformation (\ref{eq:T}) 
of Eq.~(\ref{eq:A}), i.e., $\mu(t)\equiv0\Longrightarrow c_1'(t)=r(t)$.
The latter is nothing but the Riccati equation whose solution can be 
written in terms of the logarithmic derivative of the Gau{\ss}
hypergeometric function. An explicit formula for the solution of the 
sixth Painlev\'e equation is than easy to get from the 
Eqs.~(\ref{sys:Ansc}) and a parametrization of the solution of the 
sixth Painlev\'e equation in terms of matrix elements of $A_k$
given in \cite{JM}. This solution corresponds to $\theta_t=\pm1$.

It is important to notice that there is another
construction of the one-parameter families of the solutions of the
sixth Painlev\'e equation; it is based on on the ``triangular reduction'' 
of the system (\ref{eq:A}), (\ref{eq:T})
(the latter also leads to the hypergeometric functions, but for 
$\theta_0+\theta_1+\theta_t+\theta_\infty=0$, see, e.g.,\cite{K})).
These two constructions give, modulo application of the Schlesinger 
transformations and fractional-linear transformations of $\lambda$, all 
one-parameter families of solutions of the sixth Painlev\'e equation.
The fact that there are no other solutions of the sixth Painlev\'e
equation follows from the work \cite{W}. 
It is clear that application of the non-Schlesinger deformations to
the construction of the special solutions of the 
Garnier systems and,
so-called, higher
Painlev\'e equations should ``break the symmetry'' with the construction
of the special solutions based on the triangular reduction of the
associated linear ODEs. The latter always leads to the linear 
ODEs for (generally multivariable) hypergeometric functions. Whilst
the construction related with the non-Schlesinger deformations leads to
the Riccati equations (for $2\times2$ matrix ODEs) which can be
transformed to the linear ODEs of the second order whose coefficients are
defined by solutions of the Garnier systems (Painlev\'e/higher Painlev\'e
equations) which are junior members of the corresponding hierarchies.

The most general non-Schlesinger isomonodromy deformation of Eq.~(\ref{eq:A})
are defined by the compatibility condition of Eq.~(\ref{eq:A}) with
the following ODE:
\begin{equation}
 \label{eq:TGEN}
\frac{\partial\Psi}{\partial t}=
\left(-\frac{A_t+\Lambda_1^t}{(\lambda-t)}+
\sum\limits_{k=2}^{n_t}\frac{\Lambda_k^t}{(\lambda-t)^k}+
\sum\limits_{k=1}^{n_1}\frac{\Lambda_k^1}{(\lambda-1)^k}+
\sum\limits_{k=1}^{n_0}\frac{\Lambda_k^0}{\lambda^k}\right)\Psi,
\end{equation}
where $n_t\geq1,\;n_1\geq0$, and $n_0\geq0$ are integers. 
The notation are chosen such that if $\Lambda^p_{n_p}=0$, then
$\Lambda^p_{k}=0$ for all $k\leq n_p$ and if the upper limit of the
sum is less than its lower limit, then the sum is absent.
One proves, that 
$\mathrm{tr}\,\Lambda^p_k=0$ and $\left(\Lambda^p_{n_p}\right)^2=0$ for
$p=t,\,1,\,0$. Moreover,
\begin{equation}
\Lambda^p_{n_p}\neq0\quad\Longrightarrow\quad
A_p=\frac{n_p}2\psi_0^p\sigma_3(\psi_0^p)^{-1},\quad
\Lambda^p_{n_p}=\mu^p_{l_p}(t)\psi_0^p
\left(\begin{array}{cc}
0&0\\
1&0
\end{array}\right)(\psi_0^p)^{-1},
\end{equation}
the corresponding monodromy matrix at $\lambda=p$ is $M_p=(-1)^{n_p}$, 
this means that the function $\Psi$ has the following expansion at $\lambda=p$:
\begin{equation}
  \label{eq:psi_at_p}
\Psi=\sum\limits_{k=0}^\infty\psi_k(\lambda-p)^k
(\lambda-p)^{\frac{n_p}2\sigma_3}
C,\quad \det C=\det\psi_0=1. 
\end{equation}
It follows from this expansion that the singularity at $\lambda=p$
can be removed from Eqs.~(\ref{eq:A}) and (\ref{eq:TGEN}) via
$n_p$ transformations of the type (\ref{eq:SCH}). In
general, the  non-Schlesinger isomonodromy deformation defined by  
Eq.~(\ref{eq:TGEN}) can be parametrized via $n_p+n_1+n_0$ arbitrary 
functions of $t$. More precisely: If among the ``senior'' matrices
$\Lambda_{n_k}^k,\;k=0,\,1,\,t$, only one matrix $\Lambda^p_{n_p}$
is different from $0$, then via $n_p$ Schlesinger transformations
of the form (\ref{eq:SCH}) and, possibly, (if $p\neq t$) by permutation 
of the points $0,\,1,\,t$; one transforms $\Psi$, the fundamental solution 
of the system~(\ref{eq:A}), (\ref{eq:TGEN}), to the solution
of the hypergeometric equation (\ref{eq:hyper}) $\Phi$. If any two of
the ``senior'' matrices $\Lambda^p_{n_p}\neq0$ and $\Lambda^q_{n_q}\neq0$, 
then $n_p+n_q$ Schlesinger transformations convert $\Psi$ to the
the function $\Phi=(\lambda-r)^{-(\theta_\infty+1)\sigma_3/2}$, where
$r\neq p,q$ and $r\in\{0,\,1,\,t\}$, so that
in particular $\theta_r=-\theta_\infty-1$. Finally, if all three
matrices $\Lambda^k_{n_k}\neq0,\;k=0,\,1,\,t$, then $\Psi$ can 
be presented as a multiplication of the $n_t+n_0+n_1$ Schlesinger 
transformations.

Since the consideration above is quite local this result can be generalized
for $2\times2$ matrix ODEs
\begin{equation}
 \label{eq:Agen}
\frac{d\Psi}{d\lambda}=A(\lambda)\Psi,
\end{equation}
where $=A(\lambda)\in sl_2(\mathbb{C})$ is an arbitrary rational function
of $\lambda$.\\
{\bf Definition}.
Let $t$ be a pole of $A(\lambda)$ of the first order with the residue $A_t$,
\begin{equation}
 \label{eq:B}
A(\lambda)=B(\lambda)+\frac{A_t}{\lambda-t},
\end{equation}
where $B(\lambda)$ has no pole at $\lambda=t$ as a rational function of
$\lambda$ and $A_t$ is independent of $\lambda$. 
If fundamental solution of Eq.~(\ref{eq:Agen}) satisfies also 
Eq.~(\ref{eq:T}),
then we call any solution of the system representing compatibility condition
of Eqs.~(\ref{eq:Agen}), (\ref{eq:B}), and (\ref{eq:T}), namely,
$$
\frac{\partial B(\lambda)}{\partial t}=
\left[\frac{B(\lambda)-B(t)}{\lambda-t},\,A_t\right],\quad
\frac{dA_t}{dt}=\left[B(t),A_t\right],
$$
the Schlesinger deformation of Eq.~(\ref{eq:Agen}) with respect to
the parameter $t$. 
\begin{remark}
The fourth, fifth, and sixth Painlev\'e equations can be written as 
the Schle\-singer deformations of $2\times2$ matrix linear ODEs.
\end{remark}
{\bf Definition}.
Denote $t_1$ any other {\bf first} order pole of the
rational function $A(\lambda)$, in particular, it {\bf may coincide with}
$t$. We say, that isomonodromy deformation with respect to $t$ is
{\bf $t_1$-non-Schlesinger}, iff the rational function $R(\lambda)$
in the following equation,
\begin{equation}
 \label{eq:R}
\frac{d\Psi}{dt}=\left(-\frac{A_t}{\lambda-t}+R(\lambda)\right)\Psi,
\end{equation}
has a pole at $\lambda=t_1$. 
\begin{proposition}
 \label{prop:1}
Any $t_1$-non-Schlesinger deformation of Eq.~{\rm(\ref{eq:Agen})} 
with respect to $t$ can be transformed, by a finite
number of the Schlesinger transformations of the type given by
the first equation {\rm(\ref{eq:SCH})} where $t\to t_1$ and the 
matrix $D$ is independent of $\lambda$, to the ODE of the form
\begin{equation}
 \label{eq:Awave}
\frac{d\hat\Phi}{d\lambda}=\hat A(\lambda)\hat\Phi,
\end{equation}
where the rational function 
$\hat A(\lambda)\in sl_2(\mathbb{C})$ 
has no pole at $\lambda=t_1$ whilst its other poles and their orders 
coincide with those of the function $A(\lambda)$. 

Moreover, if, additionally, the function $A(\lambda)$ is
isomonodromic with respect to $t_1$, namely,
\begin{equation}
 \label{eq:R1}
\frac{d\Psi}{dt_1}=\left(-\frac{A_{t_1}}{\lambda-t_1}+R_1(\lambda)\right)\Psi,
\end{equation}
and the set of poles {\rm(}$\in\mathbb{CP}^1${\rm)} of the rational function 
$R_1(\lambda)$ coincides with a subset {\rm(}$\in\mathbb{CP}^1${\rm)} of the 
first order poles of the function $A(\lambda)$, then, possibly by applying a
finite number of additional Schlesinger transformations of the type 
described in the previous paragraph, one arrives at
Eq.~{\rm(\ref{eq:Awave})}, where the
function $\hat A(\lambda)$ is independent of $t_1$ {\rm(}note that
if $t_1=t$, then $R(\lambda)=R_1(\lambda)$ has a pole at 
$\lambda=t${\rm)}.
The set of poles of $\hat A(\lambda)$ is a subset of the poles of $A(\lambda)$.
\end{proposition}
\begin{remark}
We call the non-Schlesinger deformations described in the second paragraph of
Proposition {\rm\ref{prop:1}} {\rm weak non-Schlesinger deformations}. If 
one allows that among the poles of $R_1(\lambda)$ there are some 
non-Fuchsian singularities of Eq.~{\rm(\ref{eq:Agen})}, then such
isomonodromy deformations could be called {\rm strong non-Schlesinger
deformations}. For the strong non-Schlesinger deformations the latter
statement of Proposition {\rm\ref{prop:1}}, in general, is not true.\\
If we suppose that Eq.~{\rm(\ref{eq:Agen})} 
suffers from some week non-Schlesinger deformation with respect to $t$
and $R(\lambda)$ {\rm(}see Eq.~{\rm(\ref{eq:R})}{\rm)} has no pole at 
$\lambda=t$, then, by applying proper Schlesinger
transformations, Eq.~{\rm(\ref{eq:Agen})} can be simplified to 
Eq.~{\rm(\ref{eq:Awave}\rm)} with the matrix $\hat A(\lambda)$ having less 
poles than $A(\lambda)$ and the Schlesinger $t$-dependence.  
\end{remark}

\section{Non-Schlesinger Isomonodromy Deformations of
$3\times3$ matrix ODEs}
 \label{sec:3}
Whilst in the case when $2\times2$ matrix ODE~(\ref{eq:Agen}) with
the connection matrix (\ref{eq:B})
suffers from the weak non-Schlesinger isomonodromy deformations 
with respect to $t$, it is always reducible via the Schlesinger 
transformations to a simpler ODE with a less number of poles, 
this is not always the case for the ODEs of higher matrix dimension. 
For example, one can take a direct sum of two $2\times2$ equations of 
the type (\ref{eq:A}):
coefficients of the first equation are deformed by the Schlesinger
deformation (\ref{eq:T}), whilst coefficients of the second one 
suffer from a non-Schlesinger deformation of the type (\ref{eq:T1}). 
Clearly, we have constructed a non-Schlesinger isomonodromy
deformation of the $4\times4$ ODE with respect to $t$ which 
(in general situation) cannot be reduced via any Schlesinger 
transformation to any equation independent of $t$. However, it can 
be transformed (via an appropriate Schlesinger transformation) to the 
equation which deformation with respect to $t$ is of the Schlesinger type.  
So, the natural question is: are there any non-Schlesinger 
isomonodromy deformations of Eq.~(\ref{eq:Agen}) which are not reducible 
via Schlesinger transformations to the Schlesinger deformations of some other
ODE of the same type (\ref{eq:Agen})?     

Consider the simplest non-Schlesinger isomonodromy deformation of 
Eqs.~(\ref{eq:Agen}), (\ref{eq:B}) of the type (\ref{eq:T1}) for 
the case of $3\times3$ matrices. Eq.~(\ref{eq:L}) now implies:
$\mathrm{tr}\,\Lambda=0$, $\Lambda^3=0$.
Using this one finds that the only possible solutions of Eq.~(\ref{eq:L})
reads,
\begin{eqnarray}
 \label{eq:3AL1}
&1.& A_t=\psi_0\left(\begin{array}{ccc}
1&0&0\\
0&0&0\\
0&0&-1
\end{array}\right)\psi_0^{-1},\quad
\Lambda=\psi_0\left(\begin{array}{ccc}
0&0&0\\
\mu_1(t)&0&0\\
0&\mu_2(t)&0
\end{array}\right)\psi_0^{-1},\\
 \label{eq:3AL2}
&2.& A_t=\psi_0\left(\begin{array}{ccc}
1/3&0&0\\
0&1/3&0\\
0&0&-2/3
\end{array}\right)\psi_0^{-1},\quad
\Lambda=\psi_0\left(\!\begin{array}{ccc}
0&0&0\\
0&0&0\\
\mu_1(t)&\mu_2(t)&0
\end{array}\!\right)\psi_0^{-1},\\
 \label{eq:3AL3}
&3.& A_t=\psi_0\left(\begin{array}{ccc}
1/3&1&0\\
0&1/3&0\\
0&0&-2/3
\end{array}\right)\psi_0^{-1},\quad
\Lambda=\psi_0\left(\begin{array}{ccc}
0&0&0\\
0&0&0\\
0&\mu_2(t)&0
\end{array}\right)\psi_0^{-1},\\
 \label{eq:3AL4}
&4.& A_t=\psi_0\left(\begin{array}{ccc}
2/3&0&0\\
0&-1/3&0\\
0&0&-1/3
\end{array}\right)\psi_0^{-1},\quad
\Lambda=\psi_0\left(\begin{array}{ccc}
0&0&0\\
\mu_1(t)&0&0\\
\mu_2(t)&0&0
\end{array}\right)\psi_0^{-1},\\
\label{eq:3AL5}
&5.& A_t=\psi_0\left(\begin{array}{ccc}
2/3&0&0\\
0&-1/3&1\\
0&0&-1/3
\end{array}\right)\psi_0^{-1},\quad
\Lambda=\psi_0\left(\begin{array}{ccc}
0&0&0\\
\mu_1(t)&0&0\\
0&0&0
\end{array}\right)\psi_0^{-1}.
\end{eqnarray}   
The matrix $\psi_0$ is a function of $t$ with $\det\psi_0=1$,
$\mu_1(t)$ and $\mu_2(t)$ are some functions of $t$. Actually, from the system
(\ref{eq:AL}) follows that for any choice of the functions $\mu_k(t)$
there is a solution of the system (\ref{eq:AL}). This means that
general solution in each case depends on two 
(the cases $1$, $2$, $4$) or one (the cases $3$, $5$) arbitrary functions 
of $t$. The functions $\mu_1(t)$ and $\mu_2(t)$ may also depend on some
other pole parameters (if any). 

Consider the first case. Any fundamental solution 
at $\lambda=t$ has the following asymptotic expansion,
\begin{eqnarray}
 \label{eq:3psi_at_t}
\Psi&\underset{\lambda\to t}=&\sum\limits_{k=0}^\infty\psi_k(\lambda-t)^k
\left(\begin{array}{ccc}
\lambda-t&0&0\\
0&1&0\\
0&0&1/(\lambda-t)
\end{array}\right)
(\lambda-t)^{\Xi}C,\\
\Xi&=&
\left(\begin{array}{ccc}
0&\kappa_2&\kappa_1\\
0&0&\kappa_3\\
0&0&0
\end{array}\right),\qquad \det C=\det\psi_0=1,
 \label{eq:xi}
\end{eqnarray}
where $\kappa_l,\;l=1,\,2,\,3$ are parameters independent of $t$.
Denote 
$$
Q=C'(t)C^{-1}(t),\quad\Rightarrow\quad\mathrm{tr}\,Q=0.
$$ 
Now using expansion (\ref{eq:3psi_at_t}) one proves that
\begin{eqnarray}
 \label{eq:3T1}
\frac{d\Psi}{dt}&=&\left(\frac{\Lambda_2}{(\lambda-t)^2}-
\frac{A_t+\Lambda}{\lambda-t}\right)\Psi,\\
\Lambda_2&=&\psi_0\left(\begin{array}{ccc}
0&0&0\\
0&0&0\\
Q_{31}&0&0\\
\end{array}\right)\psi_0^{-1},\qquad
\Lambda=-\psi_0\left(\begin{array}{ccc}
0&0&0\\
Q_{21}&0&0\\
0&Q_{32}&0\\
\end{array}\right)\psi_0^{-1},
\end{eqnarray}
where $Q_{ik}$ are matrix elements of $Q$ and $A_t$ is given by 
Eq.~(\ref{eq:3AL1}). In fact, Eq.~(\ref{eq:T1}) is a
special case of Eq.~(\ref{eq:3T1}) corresponding to the case
$Q_{31}=0$. The monodromy matrix of the fundamental 
solution with expansion (\ref{eq:3psi_at_t}) corresponding to a
small loop around $\lambda=t$ reads,
$$
M_t=C^{-1}e^{2\pi i\Xi}C.
$$
So that the isomonodromy condition, $\frac d{dt}M_t=0$ is equivalent to
\begin{equation}
 \label{eq:3iso}
\big[Q,e^{2\pi i\Xi}\big]=0.
\end{equation} 
The result of the analysis of Eq.~(\ref{eq:3iso}) can be formulated as
the following 
\begin{proposition}
 \label{31}
In generic situation, i.e., $\nu(t)=Q_{31}\neq0$ or $\nu(t)=Q_{31}=0$ but 
$\mu_1(t)\mu_2(t)=Q_{21}Q_{32}\neq0$, the isomonodromy condition
{\rm(\ref{eq:3iso})} implies $\Xi=0$; therefore, there is a 
Schlesinger transformation which transfer Eq.~{\rm(\ref{eq:Agen})} with 
the Fuchsian singularity at $\lambda=t$ into the equation of the same type 
but without this singularity.

If $\nu(t)=0,\,\mu_2(t)=0$ but $\mu_1(t)\neq0$, then $\kappa_2=0$. 
If $\nu(t)=0,\,\mu_1(t)=0$ but $\mu_2(t)\neq0$, then $\kappa_3=0$.
\end{proposition}
\begin{remark}
 \label{rem:31}
In each of the two last cases of Proposition {\rm\ref{31}} there
are some further necessary conditions on $Q$, following from 
Eq.~{\rm(\ref{eq:3iso})}, which allow to have $\Xi\neq0$. The question is
whether these conditions are sufficient, namely, can one construct
a non-Schlesinger isomonodromy deformation of the type {\rm(\ref{eq:T1})} 
for some ODE of the form {\rm(\ref{eq:Agen})} such that 
Eq.~{\rm(\ref{eq:3iso})} is satisfied and the matrix $\Xi\neq0$, is opened.
\end{remark}
The analysis of the cases 2 and 4 
(see Eq.~(\ref{eq:3AL2}) and (\ref{eq:3AL4})) essentially repeats
the previous one. Some minor modifications, say, for the case 2
are as follows. Matrix $\Lambda_2=0$, $\mu_1(t)=Q_{31}$, and 
$\mu_2(t)=Q_{32}$. 
Instead of expansion (\ref{eq:3psi_at_t}) we have
\begin{equation}
 \label{eq:3psi_at_t2}
\Psi\underset{\lambda\to t}=\sum\limits_{k=0}^\infty\psi_k(\lambda-t)^k
\left(\begin{array}{ccc}
(\lambda-t)^{\frac13}&0&0\\
0&(\lambda-t)^{\frac13}&0\\
0&0&(\lambda-t)^{-\frac23}
\end{array}\right)
(\lambda-t)^{\Xi}C,
\end{equation}
where the matrices $\Xi$ and $C$ have the same properties as those
in (\ref{eq:xi}). The monodromy matrix around $\lambda=t$ reads
$M_t=e^{2\pi i/3}C^{-1}e^{\Xi}C$, therefore isomonodromy condition
again reduces to Eq.~(\ref{eq:3iso}).   
So that the part of Proposition \ref{31} for $\nu(t)=0$
and Remark \ref{rem:31} apply also for the cases 2 and 4 without any
modifications.

Minor modifications are required also in the remaining cases 3 and 5
(see Eq.~(\ref{eq:3AL3}) and (\ref{eq:3AL5})). In the case 3
expansion of the function $\Psi$ at $\lambda=t$ is just a special
case of the expansion corresponding to the case 2 
(\ref{eq:3psi_at_t2}) in which $\kappa_1=0$ and $\kappa_2=1$
see \cite{G}. The non-Schlesinger condition $\mu_2(t)=Q_{32}\neq0$ 
implies $\kappa_3=0$. Similarly, the non-Schlesinger condition
in the case 5, i.e., $\mu_1(t)=0$,  leads to the condition
$\kappa_1=0$,  $\kappa_2=0$, and $\kappa_3=1$. In these cases  
the $\log$-terms in the corresponding asymptotic expansions
remain, so that these singularities definitely cannot be removed via
Schlesinger transformations. 

Nevertheless, in the cases 3 and 5 one can transform the non-Schlesinger
deformations (\ref{eq:T1}) to the Schlesinger form (\ref{eq:A}) via the 
Schlesinger transformations. Although, in this case one cannot reduce the 
number of poles in the transformed version of Eq.~(\ref{eq:Agen}) like it 
can be done in the other cases considered above. Actually, the idea of
this proof\footnote{For the Fuchsian systems the idea of the proof was 
communicated to me by A.~A.~Bolibruch.}
is quite simple and works for general weak non-Schlesinger 
deformations of Eq.~(\ref{eq:Agen}) in $n\times n$ matrices. 
However, to avoid long introduction of the notation, I'll explain 
it by using the redundant notation, which are already introduced.     
Suppose that Eq.~(\ref{eq:R}) defines some weak non-Schlesinger deformation
of Eq.~(\ref{eq:Agen}). Denote $\{t_1,\ldots,t_m\}$ the set of poles of 
the function $R(\lambda)$. One of them may coincide with $t$ and
simultaneously they belong to the set of the {\it first order poles} of 
Eq.~(\ref{eq:Agen}). Note, that $t_1,\ldots,t_m$ are not supposed to be
the first order poles of $R(\lambda)$. Denote, $C_{1}(t),\ldots,C_{m}(t)$ 
and $\Xi_{1},\ldots,\Xi_{m}$ the matrices defining pole expansions 
(the matrices like $C(t)$ and $\Xi$ in Eqs.~(\ref{eq:3psi_at_t})) and 
(\ref{eq:3psi_at_t2}) of the solution $\Psi$ at the points 
$t_1,\ldots,t_m$, respectively. The matrices $C_{k}(t)$ may also depend
on some other parameters, in particular, on $t_{k}$, whilst the upper
nilpotent matrices $\Xi_k$ are chosen to be constants.
The key observation is that the matrices $C_k(t)C^{-1}_k(t_0)$, where $t_0$
$\notin\{t_1,\ldots,t_m\}$
is any point chosen such that all matrices $C_k(t_0)$ are
invertible, commute with $\Xi_k$. This fact follows from the isomonodromy
condition: for any monodromy matrix $M_k(t)=M_k(t_0)$. The inverse
monodromy problem at $t=t_0$ is solvable, since it is supposed that
the matrices $C_k(t_0)$ exist, therefore, we can construct the 
Schlesinger deformation with the same monodromy data as the ones for
the function $\Psi$, but with $C_k(t)\to C_k(t_0)$.
This fact follows from the
solubility of the Cauchy problem at $t=t_0$ for the (generalized)
Schlesinger system for the coefficients of Eq.~(\ref{eq:Agen}). 
Now, denote $\Psi_S$ the solution corresponding to the
Schlesinger deformations of Eq.~(\ref{eq:Agen}). Consider 
analytic properties of $\Psi\Psi_S^{-1}$ as the function of
$\lambda\in{\cal CP}^1$ and prove (Liouville theorem) that it is a 
rational function of
$\lambda$ (here the commutativity $C_k(t)C^{-1}_k(t_0)$ with $\Xi_k$
is used). So, $\Psi$ and $\Psi_S$ are related by some Schlesinger
transformation.

It is an interesting problem to classify all cases when weakly 
non - Schlesinger Eq.~(\ref{eq:Agen}) in $n\times n$ matrices can
be reduced to a simplified equation of the same type which is either
independent or ``Schlesinger'' with respect to the corresponding parameter(s).

\section{Some Examples of non-Isomonodromy Deformations of $3\times3$
matrix ODEs}
 \label{sec:4}
It is easy to see that any function $\Psi$ which solves Eq.~(\ref{eq:Agen})
with r.-h.s. given by Eq.~(\ref{eq:B}), where $B(\lambda)$ is a rational
function of $\lambda$ holomorphic at $\lambda=t$, and depends isomonodromicaly
on the parameter $t$, satisfies  also equation
\begin{equation}
 \label{eq:U}
\frac{d\Psi}{dt}=U(\lambda,t)\Psi,
\end{equation}
where $U(\lambda,t)$ is a rational function of $\lambda$. At the same
time there are some deformations of Eq.~(\ref{eq:Agen}) which defined
by Eq.~(\ref{eq:U}) with a non-rational matrix  $U(\lambda,t)$, so that
these deformations are non-isomonodromy. One interesting example of the
non-isomonodromy deformations is considered in \cite{CA}. In that example,
non-rational contribution to the matrix $U(\lambda,t)$ is proportional to
the identity matrix and therefore reducible to isomonodromy deformations
of the Schlesinger type via $\lambda$ independent gauge transformations. Here,
we consider some simplest examples of the non-isomonodromy deformations for the
following $3\times3$ matrix ODE,
\begin{equation}
 \label{eq:3A}
\frac{d\Psi}{d\lambda}=\left(\frac{A_0}{\lambda}+\frac{A}{\lambda-t}+
\left(\begin{array}{ccc}
\theta&0&0\\
0&\theta&0\\
0&0&-2\theta
\end{array}\right)\right)\Psi,
\end{equation}
where $\theta$ is a parameter. Non-isomonodromy deformations considered below 
are
also reducible to the isomonodromy ones, however, these reductions less 
straightforward
as in the example given in \cite{CA}.\\
The first deformation is defined by the following ODE,
\begin{eqnarray}
 \label{eq:3log}
\frac{d\Psi}{dt}=\left(-\frac{A}{\lambda-t}+P\ln(\lambda-t)\right)\Psi,&&
P=\left(\begin{array}{ccc}
0&1&0\\
1&0&0\\
0&0&1
\end{array}\right).
\end{eqnarray}
Denote
$$
U=A_0+A,\qquad\Theta=\mathrm{diag}\{\theta,\theta,-2\theta\},
$$
then compatibility condition of Eqs.~(\ref{eq:3A}) and
(\ref{eq:3log}) reads,
\begin{eqnarray}
 \label{eq:3AP1}
PA_0=A_0P,\quad, PU=UP,\quad P\Theta=\Theta P,\\
A_0'=\frac 1t[U,\,A_0],\quad U'=P+[\Theta,\,U-A_0].
\label{eq:3AP2}
\end{eqnarray}
Solution of the system (\ref{eq:3AP1}) and (\ref{eq:3AP2}) 
can be written as follows,
\begin{eqnarray*}
A_0&=&\left(\begin{array}{ccc}
a_1(t)&a_1(t)+\theta_1&a_3(t)\\
a_1(t)+\theta_1&a_1(t)&a_3(t)\\
a_2(t)&a_2(t)&-2a_1(t)-\theta_2
\end{array}\right),\\
U&=&\left(\begin{array}{ccc}
\theta_4&t+\theta_5&u_3(t)\\
t+\theta_5&\theta_4&u_3(t)\\
u_2(t)&u_2(t)&t+\theta_0+\theta_4+\theta_5
\end{array}\right),
\end{eqnarray*}
where $\theta_k,\;k=0,\ldots,5$ are parameters and 
\begin{eqnarray*}
&4a_1(t)+\theta_1+\theta_2=\sqrt{\theta_3^2-8a_2(t)a_3(t)},\quad
a_3(t)=-\frac{u'(\tau)}{2\sqrt2}e^\tau,\quad
a_2(t)=\frac{v'(\tau)}{2\sqrt2}e^{-\tau},&\\
&u_3(t)=\frac{u(\tau)}{2\sqrt2}e^\tau,\quad
u_2(t)=\frac{v(\tau)}{2\sqrt2}e^{-\tau},\quad
\tau=3\theta t.&
\end{eqnarray*}
The functions $u=u(\tau)$ and $v=v(\tau)$ satisfy the following system,
\begin{eqnarray}
\tau u''&=&u\sqrt{\theta_3^2+u'v'}-(\theta_0+\tau)u',\\
\tau v''&=&v\sqrt{\theta_3^2+u'v'}+(\theta_0+\tau)v'.
\end{eqnarray}
This system can be reduced to one ODE of the second order.
To prove it let us define auxiliary variables,
$$
f=uv,\quad g=\sqrt{\theta_3^2+u'v'},\quad h=u'v-v'u. 
$$ 
One finds the following equations,
\begin{equation}
 \label{sys:P5}
\begin{array}{l} 
f'=2\tau g',\quad h'=-2(\theta_0+\tau)g',\quad \tau h'=-(\theta_0+\tau)f',\\
f'^2=h^2+4(g^2-\theta_3^2)f,\quad 2\theta_0g+f+h=\kappa,
\end{array}
\end{equation}
where $\kappa$ is the constant of integration. By excluding the function
$h$ we arrive to the following system of ODEs
\begin{equation}
 \label{sys:fg}
\begin{array}{l}
\tau\big((\tau g')'-g^2+\theta_3^2\big)+
\frac12(\theta_0+\tau)(\kappa-2\theta_0g)=\frac12(2g+\theta_0+\tau)f,\\
(\tau g')^2=\frac14(\kappa-2\theta_0g-f)^2+(g^2-\theta_3^2)f.
\end{array}
\end{equation}
Finding $f$ from the first equation and substituting it into the second
one, we get an ODE for the function $g$ which is quadratic with respect to
the second derivative. This equation can be solved in terms of the
fifth Painlev\'e transcendent: the following derivation of this fact
is due to C.~M.~Cosgrove. 

First, observe that $f$ can be eliminated to give the following 
second-order second-degree differential equation for~$g$: 
\begin{eqnarray}
&\left( \tau^2 g'' + \tau g' + 2g^3 + 3\theta_0 g^2 
     - \bigl( \kappa + 2\theta_3^{\,2} \bigr)g - \theta_0\theta_3^{\,2} 
     \right)^2 & \nonumber \\
&\qquad\qquad = \, (2g + \tau + \theta_0)^2 \left( \tau^2 (g')^2 
     + \bigl( g^2 - \theta_3^{\,2} \bigr)
     \bigl( g^2 + 2\theta_0 g - \kappa - \theta_3^{\,2} \bigr) 
     \right).&  \label{geqn}
\end{eqnarray}
Equations gauge-equivalent to (\ref{geqn}) were known to Chazy 
\cite{Ch1909} and Bureau~\cite{B1972}.  The singular integral is 
immediately apparent: if the last factor on the right-hand side is 
set to zero, then the left-hand side also vanishes and $g(\tau)$ 
becomes an elliptic function of the variable $\log \tau$.  

To get the general integral, first observe that $g''$ can be 
eliminated between the first equation in (\ref{sys:fg}) and the 
derivative of the second.  The result can be factorised as 
\begin{equation} 
\left( f' - 2\tau g' \right) \left( f + 2g^2 + 2\theta_0 g - \kappa 
     - 2\theta_3^{\,2} \right) \,=\, 0.  \label{fgeqn}
\end{equation}
The second factor yields the singular integral again.  The first 
factor, which will lead to the general integral, vanishes when 
we express $f$ and $g$ in terms of an auxiliary variable $H(\tau)$ 
according to 
\begin{equation}
f \,=\, \null - 4 (\tau H' - H) - \theta_0^{\,2}\,, \qquad 
     g \,=\, \null - 2H' - \frac{1}{2} \theta_0\,.  \label{fgH}
\end{equation}

When $f$ and $g$ are eliminated in favour of~$H$, we obtain the 
following second-order second-degree equation: 
\begin{equation}
\tau^2 (H'')^2 \,=\, \null - 4(H')^2 (\tau H' - H) + A_1 (\tau H' - H)^2 
+ A_2 (\tau H' - H) + A_3 H' + A_4\,,  \label{Heqn}
\end{equation}
where 
\begin{eqnarray*}
A_1\!\! &=&\!\! 1,  \quad
A_2 \,=\, \frac{1}{4} \bigl( 3\theta_0^{\,2} + 4\theta_3^{\,2} 
     + 2\kappa \bigr),  \quad
A_3 \,=\, \frac{1}{2} \theta_0 \bigl(\theta_0^{\,2} + \kappa \bigr),\\ 
A_4\!\! &=&\!\! \frac{1}{16} \left( \bigl(\theta_0^{\,2} + \kappa \bigr) 
     \bigl( 3\theta_0^{\,2} + \kappa \bigr) 
     + 4\theta_0^{\,2}\theta_3^{\,2} \right).  
\end{eqnarray*}
Equation (\ref{Heqn}) or a gauge-equivalent version appears in several 
references, including \cite{Ch1909,B1972,CS}.  We have presented 
it in the standard form of equation \mbox{SD-I.b} in~\cite{CS}. 
Its general solution is 
\begin{eqnarray}
H\!\!&=&\!\! \frac{1}{4w} \left( \frac{\tau w'}{w - 1} - w \right)^2 
     \,-\, \frac{1}{4} \bigl( \theta_0^{\,2} + \theta_3^{\,2} 
     + \kappa \bigr) (w - 1)\nonumber\\  
\qquad\qquad\null\!\!&+&\!\! \frac{\theta_3^{\,2}(w - 1)}{4w} 
     \,+\, \frac{\theta_0 \tau (w + 1)}{4(w - 1)} 
     \,-\, \frac{\tau^2 w}{4(w - 1)^2}\,,  \label{Hsoln}
\end{eqnarray}
where the variable $w(\tau)$ is any solution of the Painlev\'e-V 
equation, 
\begin{equation}
w''=\left( \frac{1}{2w} + \frac{1}{w-1} \right) (w')^2 
     - \frac{1}{\tau}\,w' + \frac{(w-1)^2}{\tau^2} \left( 
     \alpha w + \frac{\beta}{w} \right) + \frac{\gamma w}{\tau} 
     + \frac{\delta w(w+1)}{w-1}\,,  \label{PV}
\end{equation}
having parameters,
$$
\alpha \,=\, \frac{1}{2} \bigl( 1 + \sigma)^2, \quad 
     \mathrm{where}\quad\sigma \,:=\, \null \pm \sqrt{\theta_0^{\,2} 
     + \theta_3^{\,2} + \kappa \,}, \quad
\beta \,=\, \null - \frac{1}{2}\theta_3^{\,2}\,, \quad 
     \gamma \,=\, \theta_0\,, \quad 
     \delta \,=\, \null - \frac{1}{2}\,. 
$$
The Schlesinger and Lukashevich transformations admitted by the 
Painlev\'e-V equation induce corresponding symmetries in the 
$f,g$-system~(\ref{sys:fg}).  One of these, which is easily found 
directly, is the involution, 
\begin{eqnarray*}
f &\to& f - 2\theta_3(\theta_3 + \sigma) - \kappa,\quad  
g\,\to\, g + \frac{1}{2}(\theta_0 - \theta_3 - \sigma),\quad
\sigma \,\to\, \frac{1}{2} (\theta_0 - \theta_3 + \sigma),  \\
\theta_0 &\to& \theta_3 + \sigma, \quad
\theta_3 \,\to\, \frac{1}{2}(\theta_0 + \theta_3 - \sigma),\quad  
\kappa \, \to\, (\theta_0 - 2\theta_3)\sigma - \theta_0^{\,2} 
     - \theta_0\theta_3 - 2\theta_3^{\,2} - \kappa.
\end{eqnarray*}

As is well known, there are several classes of Painlev\'e-V 
equations that can be transformed into the Painlev\'e-III equation, 
\begin{equation}
  \label{PIII}
u'' \,=\, \frac{(u')^2}{u}\, - \frac{u'}{\tau}\, 
     + \tilde{\alpha} u^3 + \frac{\tilde{\beta}}{\tau}\,u^2 
     + \frac{\tilde{\gamma}}{\tau} + \frac{\tilde{\delta}}{u}\,.  
\end{equation}
The case $\theta_0 = \kappa = 0$, with $\theta_3$ arbitrary, is one 
such example.  In terms of an auxiliary variable $p(\tau)$, we have 
\begin{equation}
 \label{fgp}
f \,=\, p^2, \qquad g \,=\, \tau p^{-1} \bigl( p'' - \frac{1}{4}p 
     \bigr),  
\end{equation}
where $p$ satisfies the second-order second-degree equation, 
\begin{equation}
 \label{peqn}
\tau^2 \left( p'' - \frac{p}{4} \right)^2 \,=\, 
     p^2 \left( (p')^2 - \frac{p^2}{4} + \theta_3^{\,2} \right),  
\end{equation}
Equation (\ref{peqn}) 
is a special case of an equation appearing in \cite{Ch1909,B1972,CS}.  
Its solution is given by 
\begin{equation}
 \label{psoln}
p \,=\, \tau u^{-1} \bigl(u' + u^2 - \frac{1}{16} \bigr),  
\end{equation}
where $u(\tau)$ is any solution of the Painlev\'e-III equation (\ref{PIII}) 
with parameters, 
$$
\tilde{\alpha} \,=\, 1, 
     \qquad \tilde{\beta} \,=\, \null - (1 \pm 2\theta_3), 
     \qquad \tilde{\gamma} \,=\, \frac{1}{16}(1 \pm 2\theta_3), 
     \qquad \tilde{\delta} \,=\, \null - \frac{1}{256}\,.  
$$
The involution $u \to 1/(16u)$ induces the reflection $p \to -p$, which 
leaves $f$ and $g$ invariant.  The above involution admitted by $f$ 
and $g$ induces a similar Painlev\'e-III solution of the $f,g$-system 
having $\kappa = - \theta_0^{\,2}$ and \mbox{$\theta_3 = 0$}.

Of course, we can choose more general form of the matrix $P$ in these
examples, in particular it is clear that everything go through for
$P\longrightarrow\mu(t)P$ where $\mu(t)$ is an arbitrary function.
This will lead to a generalization of Eq.~(\ref{sys:fg}) which contains
the function $\mu(t)$. However, most likely the latter equation will be
again equivalent to the fifth Painlev\'e equation, since as is shown in
the work \cite{CS} the class of equations quadratic with respect to
the second derivatives and solvable in terms of the Painlev\'e transcendents
contains equations with the coefficients depending on some arbitrary 
functions which are not removable by simple scaling transformations.\\
An explanation of the appearance of the Painlev\'e transcendents in
the description of the non-isomonodromy deformations is that
these deformations preserve some subset of the monodromy data of 
Eq.~(\ref{eq:3A}). More precisely,  the following transformation,
$\Psi=Q\tilde\Psi$ where the matrix 
$$
Q=\left(\begin{array}{ccc}
0&1&1\\
0&1&-1\\
1&0&0
\end{array}\right)
$$ 
diagonalizes $P$, $Q^{-1}PQ=\mathrm{(1,1,-1)}$; transforms
system (\ref{eq:3A}), (\ref{eq:3log}) to the block form,
$$
\left(\begin{array}{ccc}
~*&*&0\\
~*&*&0\\
0&0&*
\end{array}\right).
$$
Therefore, denoting $\hat A$ and $\hat A_0$ 
$2\times2$ minors with non-vanishing elements
of the matrices $Q^{-1}AQ$ and $Q^{-1}A_0Q$
correspondingly, one finds that vector, $\hat\Psi$,
formed with the first two
components of the vector $\tilde\Psi$,
solves the following system of $2\times2$ ODEs,
\begin{equation}
 \label{sys:p5}
\frac{d\hat\Psi}{d\lambda}=\left(\left(\begin{array}{cc}
-2&0\\
0&1
\end{array}\right)+
\frac{\hat A_0}\lambda+\frac{\hat A}{\lambda-t}\right)\hat\Psi,\qquad 
\frac{d\hat\Psi}{dt}=-\frac{\hat A}{\lambda-t}\hat\Psi.
\end{equation}
As is well known from \cite{JM} the monodromy data of fundumental
solutions of system (\ref{sys:p5})
are independent of $t$ and correspoding isomonodromy deformations of
its coefficients are governed by the fifth Painlev\'e equation.
This is another derivation of the relation of the system (\ref{sys:fg}) with
the fifth Painlev\'e equation.

Another deformation of Eq.~(\ref{eq:3A}) can be defined
as follows,
\begin{eqnarray}
 \label{eq:alpha}
\frac{d\Psi}{dt}=\left(-\frac{A}{\lambda-t}+P_1(\lambda-t)^\alpha\right)\Psi,&&
P_1=\left(\begin{array}{ccc}
0&1&0\\
0&0&0\\
0&0&0
\end{array}\right).
\end{eqnarray}
The compatibility condition of (\ref{eq:alpha}) with (\ref{eq:3A}) reads
\begin{eqnarray}
 \label{sys:p11}
[\Theta,P_1]=[A_0,P_1]=0,&&[A_1,P_1]=\alpha P_1,\\
A_0'=\frac 1t[A_1,A_0],&&A_1'=\frac 1t[A_0,A_1]+[\Theta,A_1].
\label{sys:p12}
\end{eqnarray}
In the case of $3\times3$ matrices the system (\ref{sys:p11}), (\ref{sys:p12})
reduces to the equation for the confluent hypergeometyric functions,
for higher matrix dimensions it is a system of nonlinear ODEs.

One can consider more complicated than (\ref{eq:3log}) and (\ref{eq:alpha})
non-isomonodromy deformations of Eq.~(\ref{eq:3A}) by adding into the
r.-h.s.'s of these equations some further terms (like $\log^2$, etc.).
I believe that such    
non-isomonodromy deformations requires further study. In particular,
a combination of the non-Schlesinger terms, the terms considered in this 
and the previous sections, could possibly lead to non-Painlev\'e type
equations.

{\bf Acknowledgment} 
I am grateful to A.~A.~Bolibruch who called
my attention (after this work has been done) 
to his work \cite{B} and for subsequent valuable discussions, and
Nalini Joshi who pointed me out the work \cite{CA}. 
I am also indebt to C.~M.~Cosgrove for his help in the identification
of equations (\ref{sys:P5}) and (\ref{peqn}). 
I would like to thank A.~P.~Fordy and V.~B.~Kuznetsov,
the organizers of the Workshop {\it Mathematical
Methods in Regular Dynamics}, for the invitation and partial financial 
support. My participation was also partially supported by Russian 
Foundation for Basic Research and Australian Research Council.

\end{document}